\documentclass[amsmath,aps,showpacs,a4paper,10pt]{revtex4}

 \usepackage{epsf}
 \usepackage{graphicx}    
 \usepackage{enumerate}   

\begin{document}

 \newcommand{\be}[1]{\begin{equation}\label{#1}}
 \newcommand{\ee}{\end{equation}}
 \newcommand{\bea}{\begin{eqnarray}}
 \newcommand{\eea}{\end{eqnarray}}
 \def\disp{\displaystyle}

%
 \newcommand{\dmunit}{ {\rm pc\hspace{0.24em} cm^{-3}} }
 \newcommand{\cmn}{,\,}

 \begin{titlepage}

 \begin{flushright}
 arXiv:2111.07476
 \end{flushright}


 \title{\Large \bf Fast Radio Burst Distributions Consistent with\\
 the First CHIME/FRB Catalog}

 \author{Da-Chun~Qiang\,}
 \email[\,email address:\ ]{875019424@qq.com}
 \affiliation{School of Physics,
 Beijing Institute of Technology, Beijing 100081, China}

 \author{Shu-Ling~Li\,}
 \affiliation{School of Physics,
 Beijing Institute of Technology, Beijing 100081, China}

 \author{Hao~Wei\,}
 \email[\,Corresponding author;\ email address:\ ]{haowei@bit.edu.cn}
 \affiliation{School of Physics,
 Beijing Institute of Technology, Beijing 100081, China}

 \begin{abstract}\vspace{1cm}
 \centerline{\bf ABSTRACT}\vspace{2mm}
 Currently, fast radio bursts (FRBs) have become a very active field in
 astronomy and cosmology. However, the origin of FRBs is still unknown to
 date. The studies on the intrinsic FRB distributions might help us to
 reveal the possible origins of FRBs, and improve the simulations for
 FRB cosmology. Recently, the first CHIME/FRB catalog of 536 events was
 released. Such a large uniform sample of FRBs detected by a single
 telescope is very valuable to test the FRB distributions. Later, it has
 been claimed that the FRB distribution model tracking the cosmic star
 formation history~(SFH) was rejected by the first CHIME/FRB catalog. In the
 present work, we consider some empirical FRB distribution models, and find
 that many of them can be fully consistent with the CHIME/FRB observational
 data for some suitable model parameters. Notice that a suppressed evolution
 with respect to SFH is commonly found for FRBs. In particular,
 we independently confirm that the FRB distribution model tracking SFH
 can be rejected at very high confidence. On the other hand, all the
 ``\,successful\,'' models effectively require a certain degree
 of ``\,delay\,'' with respect to SFH. These results might shed light on the
 origin of FRBs and FRB cosmology.
 \end{abstract}


 \maketitle

 \end{titlepage}

 \renewcommand{\baselinestretch}{1.0}


\section{Introduction}\label{sec1}

Fast radio bursts (FRBs) are mysterious transient radio sources
 of millisecond duration~\cite{NAFRBs,Lorimer:2018rwi,Keane:2018jqo,
 Kulkarni:2018ola,Pen:2018ilo,Petroff:2019tty,Cordes:2019cmq,
 Zhang:2020qgp,Xiao:2021omr,Petroff:2021wug}. More than a decade passed
 since their discovery~\cite{Lorimer:2007qn,Thornton:2013iua}, but the
 origin of FRBs is still unknown to date~\cite{Petroff:2019tty,
 Cordes:2019cmq,Zhang:2020qgp,Xiao:2021omr,Petroff:2021wug}.
 Many theoretical models have been proposed for the engines of FRBs in
 the literature (see e.g.~\cite{Petroff:2019tty,Cordes:2019cmq,
 Zhang:2020qgp,Xiao:2021omr,Petroff:2021wug} for comprehensive reviews
 and \cite{Platts:2018hiy} for the up-to-date online catalogue of FRB
 theories). These theoretical models predicted quite different
 FRB distributions, especially the intrinsic redshift distribution for
 FRBs (see e.g. Fig.~1 of~\cite{Zhang:2020ass} for a nice summary). So,
 the studies on the intrinsic FRB distributions might help us to reveal
 the possible origins of FRBs.

The large dispersion measures (DMs) of observed FRBs well in excess of
 the Galactic values suggest that FRBs are at extragalactic/cosmological
 distances. Thus, it is reasonable to study cosmology and the intergalactic
 medium (IGM) by using FRBs (see e.g.~\cite{Deng:2013aga,Yang:2016zbm,
 Gao:2014iva,Zhou:2014yta,Yu:2017beg,Yang:2017bls,Wei:2018cgd,Li:2017mek,
 Jaroszynski:2018vgh,Madhavacheril:2019buy,Wang:2018ydd,Walters:2017afr,
 Cai:2019cfw,Li:2019klc,Wei:2019uhh,Wu:2021jyk,Er:2020qbj,Zhou:2021ygz,
 Zhao:2021jeb,Qiang:2019zrs,Qiang:2020vta,Qiang:2021bwb}). To this end,
 a large number of FRBs with identified redshifts is required. However,
 this is still not available currently. In fact, only around
 20 FRB host galaxies and their redshifts were identified to
 date~\cite{Heintz:2020}. Therefore, one has to instead use the simulated
 FRBs with mock redshifts to study cosmology in the literature (see
 e.g.~\cite{Deng:2013aga,Yang:2016zbm,Gao:2014iva,Zhou:2014yta,
 Yu:2017beg,Yang:2017bls,Wei:2018cgd,Li:2017mek,Jaroszynski:2018vgh,
 Madhavacheril:2019buy,Wang:2018ydd,Walters:2017afr,Cai:2019cfw,Li:2019klc,
 Wei:2019uhh,Wu:2021jyk,Er:2020qbj,Zhou:2021ygz,Zhao:2021jeb,Qiang:2019zrs,
 Qiang:2020vta,Qiang:2021bwb}). In the simulations, one should assume an FRB
 redshift distribution, and randomly generate mock redshifts for the
 simulated FRBs by using the assumed redshift distribution. Then, one can
 study cosmology with this mock sample of simulated FRBs. However, the
 actual redshift distribution of FRBs is still unknown to date. Thus,
 various redshift distributions for FRBs have been assumed in
 the literature. But it was found in e.g.~\cite{Qiang:2021bwb} that
 some biases might rise from these assumed FRB redshift distributions.
 Therefore, it is also important to seek the realistic redshift
 distributions from the actual FRB data, which might shed light on the
 simulations in FRB cosmology.

The important FRB distributions include the intrinsic redshift distribution
 and the intrinsic energy (or luminosity) distribution. Note that the
 specific fluence ($F_\nu$) distribution is a convolution of the redshift
 and energy distributions. It is commonly believed that the specific
 fluence distribution is insensitive to these two distributions
 \cite{Petroff:2019tty,Xiao:2021omr,Zhang:2021kdu,Lu:2019pdn,
 Vedantham:2016tjy}. In the literature, it is usually modeled as a power
 law $N(>F_\nu)\propto F_\nu^\lambda$, in particular $\lambda=-3/2$ for
 a non-evolving population in Euclidean space~\cite{Petroff:2019tty,
 Xiao:2021omr,Zhang:2021kdu,Lu:2019pdn,Vedantham:2016tjy}. On the other
 hand, the isotropic energy ($E$) distribution can be well constrained
 by the observations as a power law $dN/dE\propto E^{-\alpha}$ with
 $1.8\lesssim\alpha\lesssim 2$ roughly~\cite{Zhang:2020ass,Lu:2019pdn,
 Lu:2020nsg,Luo:2018tiy,Luo:2020wfx,Wang:2019sio}. It might
 has an exponential cutoff~\cite{Lu:2020nsg,Luo:2020wfx},
 \be{eq1}
 dN/dE\propto\left(E/E_c\right)^{-\alpha}\exp\left(-E/E_c\right)\,,
 \ee
 but the cutoff energy $E_c$ is not well constrained. Although it was
 suggested to be $E_c\sim 3\times 10^{41}\,{\rm erg}$ in
 e.g.~\cite{Luo:2020wfx,Zhang:2021kdu}, other values are still acceptable.

To date, the intrinsic FRB redshift distribution is poorly known, mainly due
 to the deficiency in FRBs with identified redshifts~\cite{Heintz:2020}. As
 a crude rule of thumb, the FRB redshift $z\sim {\rm DM}/(1000\,\dmunit)$
 \cite{Lorimer:2018rwi}. The detailed theoretical ${\rm DM}-z$ relation was
 given by e.g.~\cite{Ioka:2003fr,Inoue:2003ga,Deng:2013aga,
 Macquart:2020lln} (see also e.g.~\cite{Yang:2016zbm,Qiang:2019zrs,
 Qiang:2020vta,Qiang:2021bwb,Gao:2014iva,Zhou:2014yta,Yang:2017bls,
 Li:2019klc,Wei:2019uhh}). It has been confirmed by the recent observations
 \cite{Macquart:2020lln,James:2021jbo}. Thus, one might consider DM as a
 rough proxy of redshift $z$. In principle, one could constrain the FRB
 redshift distribution with the observed fluence, energy and DM
 distributions (see e.g.~\cite{Zhang:2020ass,James:2021oep,Chawla:2021igg,
 Zhang:2021kdu}).

For a long time, it has been speculated that the FRB distribution tracks
 the cosmic star formation history (SFH)~\cite{NAFRBs,Lorimer:2018rwi,
 Keane:2018jqo,Kulkarni:2018ola,Pen:2018ilo,Petroff:2019tty,Cordes:2019cmq,
 Zhang:2020qgp,Xiao:2021omr,Petroff:2021wug}. Recently, the landmark
 Galactic FRB 200428 associated with the young magnetar SGR 1935+2154
 \cite{Andersen:2020hvz,Bochenek:2020zxn,Lin:2020mpw,Li:2020qak} confirmed
 that at least some (if not all) FRBs originate from young magnetars. Thus,
 it is reasonable to expect that the FRB distribution is closely
 correlated with star-forming activities, as observed for the rapid
 repeaters~\cite{Tendulkar:2017vuq,Marcote:2020ljw,Niu:2021bnl}. However,
 the recently discovered repeating FRB 20200120E in a globular cluster of
 the nearby galaxy M81~\cite{Bhardwaj:2021xaa,Kirsten:2021llv,
 Nimmo:2021yob} suggested that at least some FRBs are associated with old
 stellar populations \cite{Zhang:2021kdu}. It challenges FRB models that
 invoke young magnetars formed in a core-collapse supernova as powering FRB
 emission. It was proposed in e.g.~\cite{Kirsten:2021llv} that FRB 20200120E
 is a highly magnetized neutron star formed via either accretion-induced
 collapse of a white dwarf or merger of compact stars in a binary system.
 So, a delayed FRB distribution tracking compact binary mergers is also
 reasonable (see e.g.~\cite{Zhang:2020ass,Zhang:2021kdu,Cao:2018yzp,
 Locatelli:2018anc}).

In~\cite{Zhang:2020ass}, two types of intrinsic FRB distribution models
 (tracking SFH and compact binary mergers respectively) were tested by using
 the observed samples from ASKAP and Parkes telescopes, respectively. Both
 observational samples have 27 FRB events each. It was found that these
 distribution models can be consistent with the observational data as of
 October 2020, while the compact binary merger models required
 the characteristic delay timescales of $2\sim 3$ Gyr~\cite{Zhang:2020ass}.
 In~\cite{James:2021oep}, another FRB distribution model $dN/(dt\,dV)\propto
 ({\rm SFH}(z))^n/(1+z)$ was tested by using a sample of 24 non-localized
 plus 7 localized FRBs detected by ASKAP, and 20 FRBs detected by the Parkes
 multibeam system. It was found that a non-evolving population can be ruled
 out at $99.9\%$ C.L. The FRB distribution was consistent with SFH, but FRB
 progenitors that evolve faster than SFH were also allowed as of 2020
 \cite{James:2021oep}. Note that the $n$-th power SFH model considered
 in~\cite{James:2021oep} is a phenomenological model which lacks a physical
 meaning. In~\cite{Cao:2018yzp}, it was found that the delayed
 FRB distribution tracking compact binary mergers with the characteristic
 delay timescales of a few hundreds Myr can be consistent with 22 Parkes
 FRBs as of December 2017. In~\cite{Locatelli:2018anc}, it was found that
 the FRB distribution tracking SFH can be consistent with 20 Parkes FRBs as
 of January 2016, and the FRB distribution evolving faster than SFH can be
 consistent with 23 ASKAP FRBs as of May 2018. But the delayed
 FRB distribution cannot be consistent with both samples. Note that
 these works in e.g.~\cite{Zhang:2020ass,James:2021oep,Cao:2018yzp,
 Locatelli:2018anc} used the observational samples consisting of only a few
 tens of FRB events. Thus, the constraints were fairly loose due to the
 deficiency in the number of data, and hence almost all FRB distribution
 models can be consistent with the observational data at that time.

Recently, the first CHIME/FRB catalog of 536 events (including
 474 one-off bursts and 62 repeat bursts from 18 repeaters) was released
 in June 2021~\cite{CHIMEFRB:2021srp}. Clearly, such a large uniform sample
 of FRBs detected by a single telescope is very valuable to test the FRB
 distributions. In~\cite{Chawla:2021igg}, the first CHIME/FRB catalog was
 used to study a variety of FRB distribution models, focusing on DM and
 scattering distributions. Its main attention was paid to the additional
 sources of scattering, namely the circumgalactic medium of intervening
 galaxies and the circumburst medium. On the other hand, in
 \cite{Chawla:2021igg} only the intrinsic FRB redshift distributions
 assuming a constant comoving number density or tracking SFH
 were considered. In~\cite{Zhang:2021kdu}, the intrinsic FRB redshift,
 fluence, energy, $\rm DM_E$ distributions were confronted with the first
 CHIME/FRB catalog. Four families of redshift distribution models were
 considered. The SFH model and the accumulated model were rejected with
 high significance by the Kolmogorov-Smirnov (KS) test against both the
 energy and $\rm DM_E$ distributions inferred from the observational data.
 The delayed model and the hybrid model were not rejected by the KS test
 with both the fluence and energy criteria, at the price of a very large
 lognormal delay with a central value of 10 and 13 Gyr, respectively.
 Note that these $10\sim 13$ Gyr delays are significantly larger than
 the one of short gamma-ray bursts ($2\sim 3$ Gyr)~\cite{Zhang:2020ass,
 Zhang:2021kdu}, and they are actually comparable with the entire life of
 our universe ($\sim 13.8$ Gyr)~\cite{Aghanim:2018eyx}. On the other hand,
 both the delayed model and the hybrid model were still rejected by the KS
 test with respect to the $\rm DM_E$ criterion. It was argued
 in~\cite{Zhang:2021kdu} that the value of $\rm DM_{host}$
 might partially account for the $\rm DM_E$ discrepancy.

In the present work, we try to find some empirical FRB distribution
 models fully consistent with the first CHIME/FRB catalog. We extend the
 works in~\cite{Zhang:2020ass,Zhang:2021kdu} and try to make these FRB
 distribution models can pass all the fluence, energy, $\rm DM_E$ criteria
 being not rejected by the KS test, even with the same $\rm DM_{host}$ as
 in~\cite{Zhang:2020ass,Zhang:2021kdu}. In Sec.~\ref{sec2}, we briefly
 describe the methodology following~\cite{Zhang:2020ass,Zhang:2021kdu}, but
 with some key modifications. In Sec.~\ref{sec3}, we propose our FRB
 distribution models, and confront them with the first CHIME/FRB catalog.
 We find that many of them can be fully consistent with the observational
 data for some suitable model parameters. In Sec.~\ref{sec4},
 some brief concluding remarks are given.


\section{Methodology}\label{sec2}

At first, we briefly describe the methodology used in the present work,
 following~\cite{Zhang:2020ass,Zhang:2021kdu} but with some key
 modifications making sensible differences. The key idea is to confront
 the Monte Carlo simulations with the observational data. If the simulations
 are rejected by the observational data, the assumed FRB distribution models
 generating these simulations could be ruled out. Otherwise, they survive.


\subsection{Monte Carlo simulations}\label{sec2a}

As in~\cite{Zhang:2020ass,Zhang:2021kdu}, we perform the Monte Carlo
 simulations in two steps: generating mock FRBs from the assumed
 intrinsic FRB distribution models, and then filtering them due to the
 telescope's sensitivity threshold and instrumental selection effects
 near the threshold. The filtered FRBs as a mock ``\,observed\,'' sample
 will be confronted with the observational data.

First, we generate mock FRBs characterized by the isotropic energy $E$,
 redshift $z$, specific fluence~$F_\nu$, and $\rm DM_E$ (the extragalactic
 DM). Following~\cite{Zhang:2020ass,Zhang:2021kdu}, we adopt the energy
 distribution modeled by Eq.~(\ref{eq1}). In principle, $\alpha$ and $E_c$
 are both free model parameters as in~\cite{Zhang:2020ass,Zhang:2021kdu}. To
 save the computational power and time, it is better to reduce the number of
 free model parameters. One can see that they take almost the same values
 for all the four FRB distribution models in \cite{Zhang:2021kdu}. Thus, we
 fix $\alpha=1.9$ and $\log E_c=41$ in the present work, since they hold for
 four and three models in~\cite{Zhang:2021kdu} respectively, where the
 energy is in units of erg, and ``\,$\log$\,'' gives the logarithm to base
 10. On the other hand, it is useless to simulate very low-energy FRBs at
 cosmological distances, because they cannot be detected by the telescope
 with a threshold in specific fluence. We find that the lower bound
 $10^{37}\,{\rm erg}$ is suitable for the CHIME telescope under
 consideration (actually we have tested other much lower bounds and found
 almost no difference after filter). Due to the exponential cutoff in
 Eq.~(\ref{eq1}), very high-energy FRBs are extremely rare. So, it is enough
 to randomly sample $E$ from the energy distribution~(\ref{eq1}) in the
 range of $10^{37}\sim 10^{44}\,{\rm erg}$.

The observed FRB redshift rate distribution reads~\cite{Zhang:2020ass,
 Zhang:2021kdu}
 \be{eq2}
 \frac{dN}{dt_{\rm obs}\,dz}=\frac{1}{1+z}\cdot\frac{dN}{dt\,dV}\cdot
 \frac{dV}{dz}\,,
 \ee
 where we have used $dt/dt_{\rm obs}=(1+z)^{-1}$ due to the
 cosmic expansion. The redshift-dependent specific comoving volume in a
 flat Friedmann-Robertson-Walker~(FRW) universe is given by
 \be{eq3}
 \frac{dV}{dz}=4\pi d_C^2\hspace{0.08em}\frac{c}{H(z)}\,,
 \ee
 where the comoving distance $d_C=d_L/(1+z)$, $d_L$ is the luminosity
 distance at redshift $z$, $c$ is the speed of light, and $H(z)$ is the
 Hubble parameter. Throughout this work, we consider the fiducial cosmology,
 namely the flat $\Lambda$CDM model, and hence
 \be{eq4}
 d_C=\frac{d_L}{1+z}=c\int_0^z\frac{d\tilde{z}}{H(\tilde{z})}\,,\quad\quad
 H(z)=H_0 E(z)=H_0\left[\Omega_m\left(1+z\right)^3+\left(1-\Omega_m\right)
 \right]^{1/2}\,,
 \ee
 where $H_0$ is the Hubble constant, and $\Omega_m$ is the
 present fractional density of dust matter (see below for their values). To
 date, the intrinsic event rate density distribution $dN/(dt\,dV)$ is poorly
 known. If the intrinsic FRB redshift distribution tracks SFH, one has
 $dN/(dt\,dV)\propto {\rm SFH}(z)$. In~\cite{Zhang:2020ass,Zhang:2021kdu},
 the analytical three-segment empirical model of~\cite{Yuksel:2008cu} was
 used for ${\rm SFH}(z)$ (see Eq.~(5) of~\cite{Zhang:2020ass}). However, the
 analytical two-segment empirical model in the well-known review on
 SFH~\cite{Madau:2014bja} (i.e. its Eq.~(15)) was used extensively in the
 literature. In this work, we instead use its updated version given
 by Eq.~(1) of~\cite{Madau:2016jbv}, namely
 \be{eq5}
 {\rm SFH}(z)\propto\frac{(1+z)^{2.6}}{1+\left((1+z)/3.2\right)^{6.2}}\,,
 \ee
 which behaves like $(1+z)^{2.6}$ and $(1+z)^{-3.6}$ at low-
 and high-redshifts, respectively. In fact, the actual $dN/(dt\,dV)$ is
 still unclear to date, and many models have been proposed for it in the
 literature. Once $dN/(dt\,dV)$ is given, we can randomly sample redshift
 $z$ from Eq.~(\ref{eq2}) for each mock FRB.

As is well known, the observed DM of FRB can be separated into
 \cite{Deng:2013aga,Yang:2016zbm,Qiang:2019zrs,Qiang:2020vta,Qiang:2021bwb,
 Gao:2014iva,Zhou:2014yta,Yang:2017bls,Li:2019klc,Wei:2019uhh}
 \be{eq6}
 {\rm DM_{obs}=DM_{MW}+DM_{halo}+DM_{IGM}+DM_{host}}/(1+z)\,,
 \ee
 where $\rm DM_{MW}$, $\rm DM_{halo}$, $\rm DM_{IGM}$, and $\rm DM_{host}$
 are the contributions from the Milky Way, the Milky Way halo,
 the intergalactic medium (IGM), and the host galaxy (including interstellar
 medium of the host galaxy and the near-source plasma), respectively. It is
 convenient to introduce the extragalactic DM as
 \be{eq7}
 {\rm DM_E=DM_{obs}-DM_{MW}-DM_{halo}=DM_{IGM}+DM_{host}}/(1+z)\,.
 \ee
 Following~\cite{Zhang:2020ass,Zhang:2021kdu}, here we adopt $\rm DM_{halo}
 =30\,\dmunit$ (see e.g.~\cite{Dolag:2014bca,Prochaska:2019})
 and $\rm DM_{host}=107\,\dmunit$ (see e.g.~\cite{Li:2020qei}). For
 a mock FRB, we can calculate the corresponding ${\rm DM_E=DM_{IGM}
 +DM_{host}}/(1+z)$ with (see e.g.~\cite{Deng:2013aga,Yang:2016zbm,
 Li:2019klc,Wei:2019uhh,Qiang:2019zrs,Qiang:2020vta,Qiang:2021bwb})
 \be{eq8}
 {\rm DM_{IGM}}=\frac{3cH_0\Omega_{b}}{8\pi G m_p}
 \int_0^z\frac{f_{\rm IGM}(\tilde{z})\,f_e(\tilde{z})\left(1+
 \tilde{z}\right)d\tilde{z}}{E(\tilde{z})}\,,
 \ee
 where $\Omega_b$ is the present fractional density of baryons, $G$ is
 the gravitational constant, $m_p$ is the mass of proton, $E(z)\equiv
 H(z)/H_0$ is the dimensionless Hubble parameter, $f_e(z)$ is the ionized
 electron number fraction per baryon, and $f_{\rm IGM}(z)$ is the fraction
 of baryon mass in IGM. The latter two are functions of redshift $z$
 in principle \cite{Li:2019klc,Wei:2019uhh,Qiang:2020vta}. Following
 \cite{Zhang:2020ass,Zhang:2021kdu}, we use the fiducial values $f_e=7/8$
 and $f_{\rm IGM}=0.84$ in this work. Unlike~\cite{Zhang:2021kdu} using
 the cosmological parameters from the Planck 2015 results
 \cite{Planck:2015fie}, we adopt $\Omega_m=0.3153$, $\Omega_b=0.0493$,
 and $H_0=67.36\,{\rm km/s/Mpc}$ instead from the Planck 2018
 results~\cite{Aghanim:2018eyx}.

Assuming a flat radio spectrum, the ``\,observed\,'' specific fluence
 for a mock FRB with isotropic energy $E$ and redshift $z$ is
 given by~\cite{Zhang:2020ass,Zhang:2018csb}
 \be{eq9}
 F_\nu=\frac{\left(1+z\right)E}{4\pi d_L^2 \nu_c}\,,
 \ee
 where $\nu_c$ is the central observing frequency, and the luminosity
 distance $d_L$ is given by Eq.~(\ref{eq4}). Note that the bandwidth
 $\Delta\nu$ usually takes the place of $\nu_c$ in the literature. But it
 was argued in~\cite{Zhang:2018csb} that it is more appropriate to use
 $\nu_c$ instead. To keep pace with~\cite{Zhang:2020ass,Zhang:2021kdu},
 we adopt the same form of $F_\nu$ in Eq.~(\ref{eq9}), with $\nu_c=600\,
 {\rm MHz}$ for CHIME~\cite{CHIMEFRB:2021srp}.

However, the mock FRBs intrinsically generated above are not the ones
 ``\,detected\,'' by the telescope, due to the telescope's sensitivity
 threshold and instrumental selection effects near the threshold. So,
 the second step is to filter them by using the telescope's sensitivity
 model, which is difficult to characterize in fact. One might use a
 sophisticated but fairly complicated model for the selection effects
 of CHIME as in~\cite{CHIMEFRB:2021srp,Chawla:2021igg}. Instead, it is
 convenient to consider the simplified sensitivity model for CHIME as
 in~\cite{Zhang:2021kdu}. Every telescope has a sensitivity threshold.
 Following~\cite{Zhang:2021kdu}, it is about $0.3\,{\rm Jy\,ms}$ for
 CHIME, or equivalently $\log F_{\nu\cmn {\rm min}}=-0.5$, where the
 specific fluence is in units of Jy\,ms. On the other hand, due to the
 direction-dependent sensitivity of the telescope, there is a ``\,gray
 zone\,'' in the $\log F_\nu$ distribution, within which CHIME has not
 reached full sensitivity to all sources~\cite{Zhang:2021kdu}. For this
 ``\,gray zone\,'', following~\cite{Zhang:2021kdu}, we adopt $\log F_{\nu
 \cmn {\rm th}}^{\rm min}=-0.5$ as its minimum threshold specific fluence,
 while its maximum threshold specific fluence $\log F_{\nu\cmn {\rm
 th}}^{\rm max}$ should be adjusted to match the ``\,observation\,''. The
 detection efficiency parameter in the ``\,gray zone\,'' is given by
 $\eta_{\rm det}={\cal R}^3$, where ${\cal R}=(\log F_{\nu\cmn {\rm th}}
 -\log F_{\nu\cmn {\rm th}}^{\rm min})/(\log F_{\nu\cmn {\rm th}}^{\rm max}
 -\log F_{\nu\cmn {\rm th}}^{\rm min})$, such that $\eta_{\rm det}\to 0$
 at $\log F_{\nu\cmn {\rm th}}^{\rm min}$ and $\eta_{\rm det}\to 1$ at
 $\log F_{\nu\cmn {\rm th}}^{\rm max}$ \cite{Zhang:2021kdu}. Outside the
 ``\,gray zone\,'' ($\log F_{\nu\cmn {\rm th}}>\log F_{\nu\cmn {\rm
 th}}^{\rm max}$), $\eta_{\rm det}=1$. We first screen out the mock FRBs
 with specific fluences below the threshold $\log F_{\nu\cmn {\rm min}}=
 -0.5$, and then filter the rest with the detection efficiency parameter
 $\eta_{\rm det}\,$. Finally, a mock ``\,observed\,'' sample of FRBs is on
 hand, which will be confronted with the observational data.


\subsection{CHIME/FRB data}\label{sec2b}

The data for the first CHIME/FRB catalog in machine-readable format can be
 downloaded from the public webpage given by~\cite{CHIMEFRB:2021srp}. It
 contains 536 events, including 474 one-off bursts and 62 repeat bursts
 from 18 repeaters. At first, we exclude the six bursts with zero fluences
 (detected directly after a system restart) as mentioned at the end of
 Sec.~3 of~\cite{CHIMEFRB:2021srp}. Then, we also exclude the bursts
 labeled with $\tt excluded_{-}flag=1$ (detected either during
 pre-commissioning, epochs of low-sensitivity, or on days with software
 upgrades) as mentioned in the middle of Sec.~6.1 and at the end of
 Table~2 of~\cite{CHIMEFRB:2021srp} (note that they might not be excluded
 in~\cite{Zhang:2021kdu}). Following~\cite{Zhang:2021kdu}, we only use the
 first detected burst of each repeating FRB and the non-repeaters. In
 practice, we identify the non-repeaters labeled with $\tt
 repeater_{-}name=-9999$ and then only take the ones labeled with $\tt
 sub_{-}num=0$ (434 bursts in total). We identify the repeaters labeled with
 $\tt repeater_{-}name\not=-9999$ and only take the ones labeled with $\tt
 sub_{-}num=0$, and then from them we adopt the first ones in each group
 with the same $\tt repeater_{-}name$ (18 bursts in total). Finally, we
 obtain an observational sample of 452 FRBs.

For each burst, its observed $F_\nu$ is given by the column labeled with
 ``\,{\tt fluence}\,'' in the data table. Its $\rm DM_E=DM_{obs}-DM_{MW}
 -DM_{halo}$, in which $\rm DM_{obs}$ is given by the column labeled with
 ``\,$\tt bonsai_{-}dm$\,'' in the data table, $\rm DM_{MW}$ is obtained by
 using NE2001~\cite{Cordes:2002wz,Cordes:2003ik}, and $\rm DM_{halo} =30\,
 \dmunit$ as mentioned above. Then, its inferred redshift $z$ is obtained
 by numerically solving ${\rm DM_E=DM_{IGM}+DM_{host}}/(1+z)$ with $\rm
 DM_{IGM}$ given by Eq.~(\ref{eq8}) and $\rm DM_{host}=107\,\dmunit$ as
 mentioned above. So, its isotropic energy $E$ can be inferred from
 Eq.~(\ref{eq9}) with the observed $F_\nu$ and the luminosity distance
 $d_L$ at redshift $z$ given by Eq.~(\ref{eq4}). Finally, the observational
 data of 452 FRBs are ready for confrontation with the mock ``\,observed\,''
 sample of FRBs.


\subsection{Strategy}\label{sec2c}

As is well known, the Kolmogorov-Smirnov (KS) test is one of the useful
 tools to compare a sample with a reference probability distribution, or to
 compare two samples~\cite{KStest}. The KS statistic quantifies the largest
 distance between the two cumulative distribution functions (CDFs) of the
 sample and the reference distribution, or two samples. One can perform
 the KS test by using {\tt scipy.stats.kstest} in Python~\cite{KStestpy},
 which returns the KS statistic and the corresponding p-value. Here we use
 the p-value ($p_{_{\rm KS}}$) in the two-sample case as in
 \cite{CHIMEFRB:2021srp,Chawla:2021igg}, rather than the KS statistic
 ($D_{\rm KS}$) used in~\cite{Zhang:2020ass,Zhang:2021kdu}. The null
 hypothesis (namely two samples are drawn from the same distribution) can
 be rejected at $90\%$ ($95\%$) confidence if $p_{_{\rm KS}}<0.1$ ($0.05$),
 respectively. Otherwise, two samples can be consistent with each other if
 $p_{_{\rm KS}}>0.1$ (or $0.05$). For two closer samples, $p_{_{\rm KS}}$
 is higher (and $p_{_{\rm KS}}=1$ for two same samples).

Following~\cite{Zhang:2020ass,Zhang:2021kdu}, the KS test can be used to
 compare the mock ``\,observed\,'' FRB samples with the CHIME/FRB samples
 of the specific fluence $F_\nu$, the isotropic energy $E$, and the
 extragalactic dispersion measure $\rm DM_E$, respectively. As mentioned
 above, the specific fluence ($F_\nu$) distribution is a convolution
 of the redshift and energy distributions, and hence the $\log N(>F_\nu)
 -\log F_\nu$ distribution is insensitive to these two distributions
 \cite{Petroff:2019tty,Xiao:2021omr,Zhang:2021kdu,Lu:2019pdn,
 Vedantham:2016tjy}. So, the following strategy was used
 in~\cite{Zhang:2021kdu}: For each redshift distribution, one can adjust
 the $E$ distribution model ($\alpha$ and $E_c$) and the sensitivity
 model ($\log F_{\nu\cmn {\rm th}}^{\rm max}$) to make the $\log F_\nu$
 distribution of the mock ``\,observed\,'' sample being not rejected by the
 KS test against the observational $\log F_\nu$ distribution. Once $\alpha$,
 $E_c\,$, and $\log F_{\nu\cmn {\rm th}}^{\rm max}$ are determined by the
 $\log F_\nu$ criterion, one can then evaluate the $\log E$ and $\rm DM_E$
 distribution criteria. The redshift distribution model is ruled out if
 the same mock FRB sample fails both $\log E$ and $\rm DM_E$
 criteria~\cite{Zhang:2021kdu}.

The above strategy of~\cite{Zhang:2021kdu} is efficient in the case of
 unfixed $E$ distribution model, since adjusting the free parameters
 $\alpha$, $E_c$ and $\log F_{\nu\cmn {\rm th}}^{\rm max}$ consumes a large
 amount of computational power and time. But a potential risk exists in
 this strategy. The $\log F_\nu$ criterion has a superior position. If one
 tends to find a fairly high $p_{_{\rm KS}}$ (or a fairly low $D_{\rm KS}$)
 for the $\log F_\nu$ criterion (say, $p_{_{\rm KS}}>0.9$ or $0.8$), the
 cases with three $p_{_{\rm KS}}$ being all larger than (say) $0.5$ or $0.4$
 for the $\log F_\nu$, $\log E$, $\rm DM_E$ criteria might be ignored.
 Many acceptable cases with three not so high but still enough $p_{_{\rm
 KS}}$ might not be found in this strategy.

In the present work, we use a different strategy. As mentioned at the
 beginning of Sec.~\ref{sec2a}, we fix the $E$ distribution model with
 $\alpha=1.9$ and $\log E_c=41$, in order to save the computational power
 and time. So, we can simultaneously scan the parameters in the redshift
 distribution model and the sensitivity model ($\log F_{\nu\cmn {\rm
 th}}^{\rm max}$) to evaluate all the three $p_{_{\rm KS}}$ for the $\log
 F_\nu$, $\log E$ and $\rm DM_E$ criteria. In our strategy, the $\log
 F_\nu$ criterion has no superior position. In this way, we might find some
 acceptable cases with three high enough $p_{_{\rm KS}}$ (say, all $>0.5$)
 for the $\log F_\nu$, $\log E$ and $\rm DM_E$ criteria at the same time.


 \begin{center}
 \begin{figure}[tb]
 \centering
 \vspace{-11mm}  
 \includegraphics[width=0.82\textwidth]{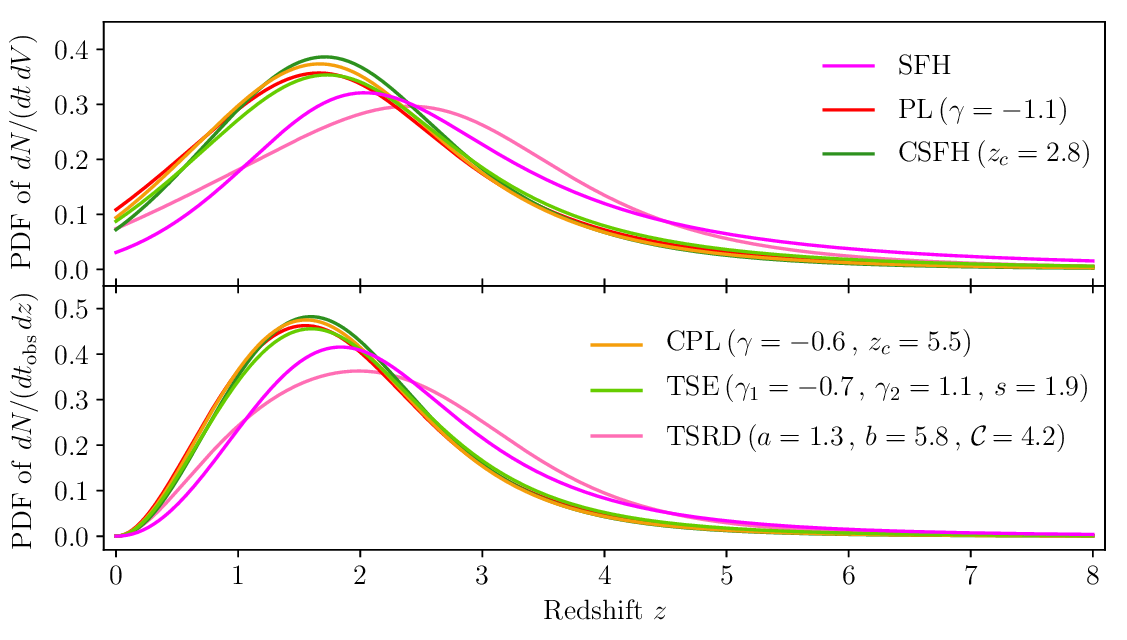}
 \caption{\label{fig1} Normalized probability distribution functions
 (PDFs) of the intrinsic FRB event rate density $dN/(dt\,dV)$ (upper
 panel) and the observed FRB event rate redshift distribution
 $dN/(dt_{\rm obs}\,dV)$ (lower panel) for various models (color online)
 with some demonstrative model parameters. Notice that they are all model
 curves predicted by Eqs.~(\ref{eq10}) or (\ref{eq14}) for $dN/(dt\,dV)$
 and Eq.~(\ref{eq2}) for $dN/(dt_{\rm obs}\,dV)$. No sensitivity threshold
 and instrumental selection effects have been applied in
 plotting Fig.~\ref{fig1}. See Sec.~\ref{sec3a} for details.}
 \end{figure}
 \end{center}


\vspace{-11mm} 


\section{Various FRB distributions versus
 the first CHIME/FRB catalog}\label{sec3}


\subsection{Various empirical FRB redshift distribution models}\label{sec3a}

History always repeats itself. In the reviews e.g.~\cite{Kulkarni:2018ola,
 Zhang:2020qgp}, it was argued that FRBs might share a similar research
 history with gamma-ray bursts (GRBs). In~\cite{Zhang:2021kdu}, it is found
 that CHIME/FRB population does not track SFH. This also happened in the
 field of GRBs many years ago~\cite{Kistler:2007ud}. The enhanced evolution
 in the GRB rate can be parameterized as $dN/(dt\,dV)\propto {\cal E}(z)
 \cdot {\rm SFH}(z)$ \cite{Kistler:2007ud} (see also e.g.~\cite{Yuksel:2008cu})
 with ${\cal E}(z)\propto (1+z)^{1.5}$. This motivates us to also consider
 some SFH-based redshift distribution models for FRBs, namely
 \be{eq10}
 dN/(dt\,dV)\propto {\cal E}(z)\cdot {\rm SFH}(z)\,,
 \ee
 where ${\rm SFH}(z)$ is given by Eq.~(\ref{eq5}). Note that these models in
 Eq.~(\ref{eq10}) with various ${\cal E}(z)$ considered in the present
 work are all phenomenological models. ${\cal E}(z)$ characterizes the
 deviation from SFH. They are motivated by the case of GRBs, but just in the
 functional form, rather than the physical meaning. Although ${\cal E}(z)
 \propto (1+z)^{1.5}$ in the case of GRBs indicates an enhanced evolution,
 we stress that ${\cal E}(z)$ in the case of FRBs can be fairly different in
 fact, which will be determined only by the observational data. Actually, a
 suppressed evolution cannot be excluded in the case of FRBs. In this work,
 we consider the following models in detail:
 \begin{itemize}
   \item SFH model, ${\cal E}(z)\equiv 1$\,;

   \item Power law ${\cal E}(z)$ (PL) model, namely
   \be{eq11}
   {\cal E}(z)\propto (1+z)^\gamma\,,
   \ee
   which reduces to the SFH model if $\gamma=0$\,;

   \item Power law ${\cal E}(z)$ with an exponential cutoff (CPL) model, namely
   \be{eq12}
   {\cal E}(z)\propto (1+z)^\gamma \exp\left(-z/z_c\right)\,;
   \ee

   \item SFH with an exponential cutoff (CSFH) model, i.e. the special case
   of CPL model with $\gamma=0\,$;

   \item Two-segment ${\cal E}(z)$ (TSE) model, namely
   \be{eq13}
   {\cal E}(z)\propto\frac{(1+z)^{\gamma_1}}{1+\left((1+z)/s
   \right)^{\gamma_1+\gamma_2}}\,,
   \ee
   which is expected to behave like $(1+z)^{\gamma_1}$ and
   $(1+z)^{-\gamma_2}$ at low- and high-redshifts respectively, if $s$
   is much larger than $1$ and $\gamma_1+\gamma_2$ is positive.
 \end{itemize}

In fact, ${\rm SFH}(z)$ given by Eq.~(\ref{eq5})
 also behaves like $(1+z)^{2.6}$ and $(1+z)^{-3.6}$ at low- and
 high-redshifts, respectively. In the PL and CPL models, the two segments of
 SFH are changed in the same way. In the TSE model, they are changed
 in different ways. This inspires us to directly consider
 a non-SFH-based model at the same level of Eq.~(\ref{eq10}), namely
 \begin{itemize}
   \item Two-segment redshift distribution (TSRD) model,
   \be{eq14}
   dN/(dt\,dV)\propto\frac{(1+z)^a}{1+\left((1+z)/{\cal C}
   \right)^{a+b\,}}\,,
   \ee
   which cannot be confused with the TSE model. It is expected to behave
   like $(1+z)^a$ and $(1+z)^{-b}$ at low- and high-redshifts respectively,
   if $\cal C$ is much larger than $1$ and $a+b$ is positive.
 \end{itemize}

In Fig.~\ref{fig1}, we show the normalized probability distribution
 functions (PDFs) of the intrinsic FRB event rate density $dN/(dt\,dV)$
 and the observed FRB event rate redshift distribution $dN/(dt_{\rm
 obs}\,dV)$ for various models with some demonstrative model parameters.
 Notice that they are all model curves predicted by Eqs.~(\ref{eq10}) or
 (\ref{eq14}) for $dN/(dt\,dV)$ and Eq.~(\ref{eq2}) for $dN/(dt_{\rm
 obs}\,dV)$. No sensitivity threshold and instrumental selection effects
 have been applied in plotting Fig.~\ref{fig1}. Clearly, all models have
 two-segment shapes, and peak around redshift $1.5\sim 2.5$. However, if
 we further consider the sensitivity threshold and instrumental selection
 effects mentioned at the end of Sec.~\ref{sec2a}, these distributions
 will be changed accordingly.


 \begin{center}
 \begin{figure}[tb]
 \centering
 \vspace{-10mm}  
 \includegraphics[width=0.82\textwidth]{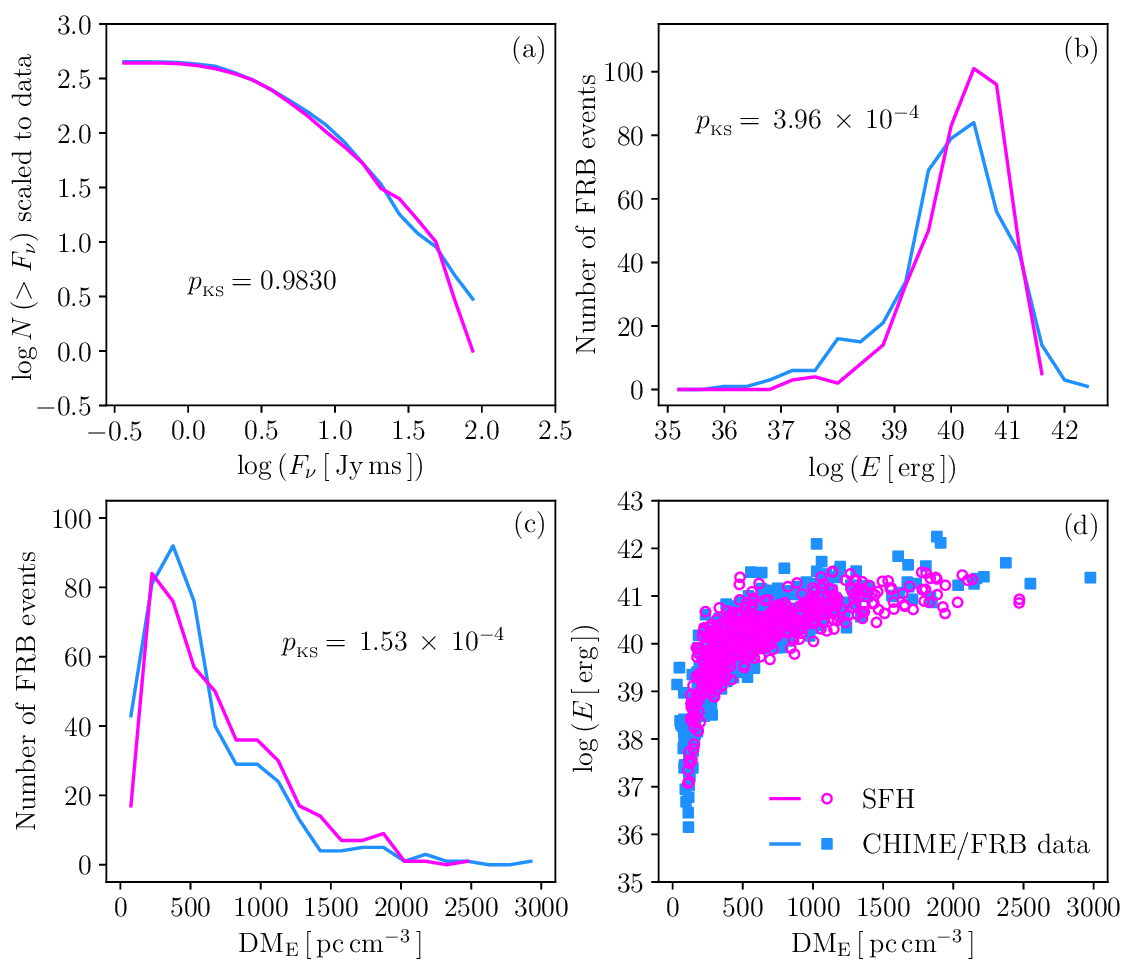}
 \caption{\label{fig2} Testing the SFH model against the CHIME/FRB data
 (color online), with respect to the (a)~$\log F_\nu$, (b)~$\log E$ and
 (c)~$\rm DM_E$ criteria. In panels (a), (b) and (c), the simulations
 are scaled to the CHIME/FRB data. The p-values $p_{_{\rm KS}}$ for the
 KS tests are given in the corresponding panels, respectively.
 In panel~(d), the ${\rm DM_E}-\log E$ distribution is also presented
 for illustration (not for test). See the text for details.}
 \end{figure}
 \end{center}


\vspace{-11mm} 


\subsection{Results}\label{sec3b}

Here, we test the six FRB redshift distribution models proposed
 in Sec.~\ref{sec3a} against the CHIME/FRB observational data. We perform
 the Monte Carlo simulations following the description given in
 Sec.~\ref{sec2a}, and filter the mock FRBs with the sensitivity threshold
 and instrumental selection effects to obtain a mock ``\,observed\,'' sample
 of FRBs as mentioned at the end of Sec.~\ref{sec2a}. In our strategy, we
 simultaneously scan the free parameters in the redshift distribution model,
 within some suitable value ranges for these parameters, as mentioned in
 Sec.~\ref{sec2c}. For each value combination of the free parameters in the
 redshift distribution model, we generate 2,000,000 (or more if necessary)
 mock FRBs to ensure that there are still enough FRBs filtered by the
 sensitivity threshold and instrumental selection effects (say, $N>300$),
 comparable with the CHIME/FRB sample consisting of 452 FRBs (and we
 strongly refer to Sec.~\ref{sec4} for the discussion about the computation
 cost). The KS tests are performed between the mock ``\,observed\,'' sample
 of FRBs and the CHIME/FRB observational sample, with respect to the $\log
 F_\nu$, $\log E$ and $\rm DM_E$ distributions. In the present work, for the
 CHIME/FRB data, we choose 20 bins in $\log F_\nu$ covering a range from
 $-0.5$ to $2.0$, in $\log E$ covering a range from $35$ to $43$, in
 $\rm DM_E$ covering a range from $0$ to $3000\;\dmunit$, respectively. For
 the simulations, if necessary, we add bins with the same bin size.


\begin{table}[tb]
 \renewcommand{\arraystretch}{1.7}
 \begin{center}
 \vspace{-8mm}  
 \begin{tabular}{ccccc} \hline\hline
 $\gamma$ &  \quad $\log F_{\nu\cmn {\rm th}}^{\rm max}$ \quad
 & \quad $p_{_{\rm KS}}$ for $\log F_\nu$ \quad
 & \quad $p_{_{\rm KS}}$ for $\log E$ \quad
 & \quad $p_{_{\rm KS}}$ for $\rm DM_E$ \quad \\ \hline
 \ \ $\bf -1.1$ \ \ & {\bf 0.72}
 & {\bf 0.9478} & {\bf 0.4905} & {\bf 0.5548} \\ \hline
 $-0.9$ & 0.75 & 0.8697 & 0.2179 & 0.3280 \\ \hline
 $-0.8$ & 0.76 & 0.8468 & 0.2257 & 0.5889 \\ \hline
 $-0.7$ & 0.68 & 0.8733 & 0.2333 & 0.3671 \\ \hline
 \hline
 \end{tabular}
 \end{center}
 \vspace{-2mm}  
 \caption{\label{tab1} Some examples of the acceptable PL models with
 the redshift distribution model parameter $\gamma$ and the sensitivity
 model parameter $\log F_{\nu\cmn {\rm th}}^{\rm max}$, as well as three
 $p_{_{\rm KS}}$ for the $\log F_\nu$, $\log E$ and $\rm DM_E$ criteria
 against the CHIME/FRB data. The boldfaced ones are also presented in
 the accompanying plots. See Sec.~\ref{sec3b} for details.}
 \end{table}



 \begin{center}
 \begin{figure}[tb]
 \centering
 \vspace{1mm}  
 \includegraphics[width=0.82\textwidth]{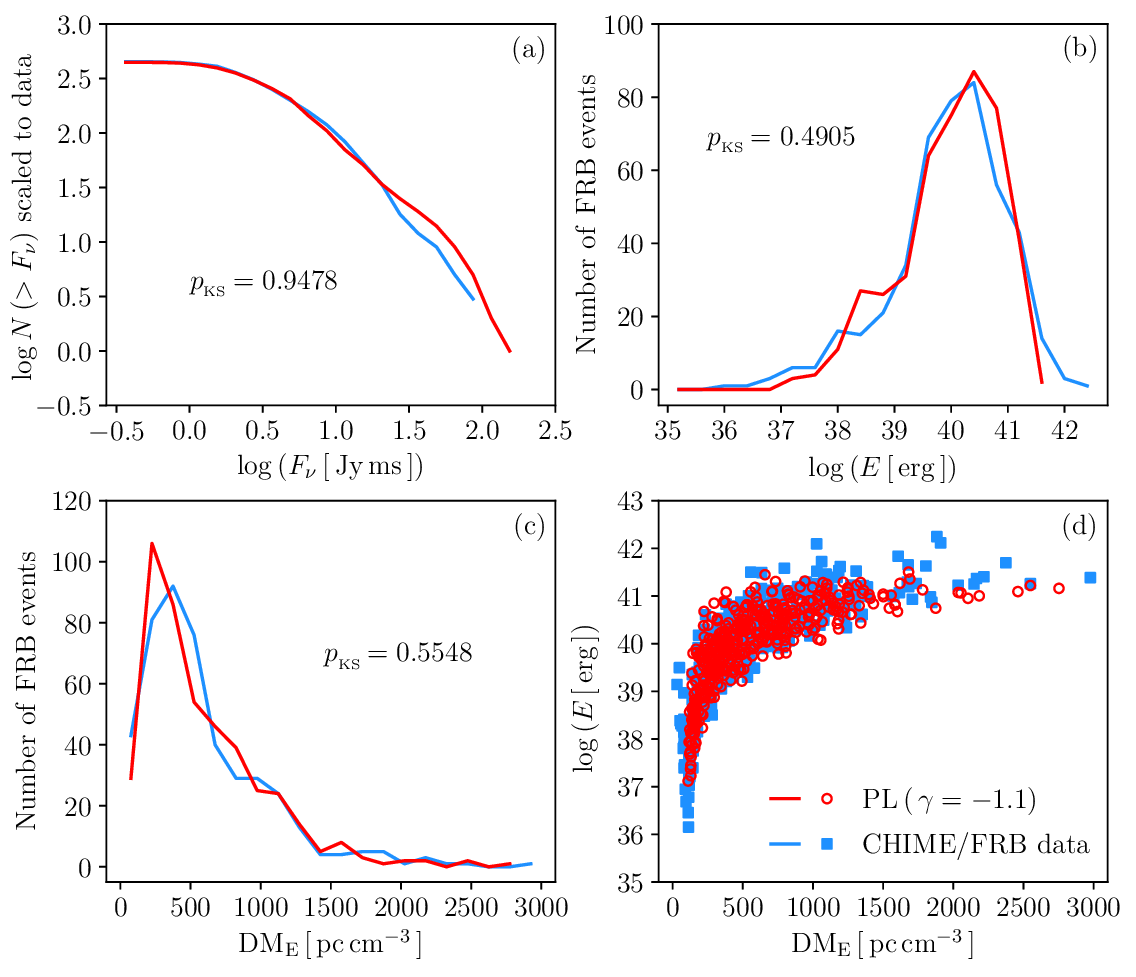}
 \vspace{-2mm}  
 \caption{\label{fig3} The same as in Fig.~\ref{fig2}, but for
 the PL model with the model parameter given in panel~(d).}
 \end{figure}
 \end{center}



\begin{table}[tb]
 \renewcommand{\arraystretch}{1.68}
 \begin{center}
 \vspace{-10mm}  
 \begin{tabular}{cccccc} \hline\hline
 $\gamma$ & \quad\ $z_c$ \quad\ & \quad $\log F_{\nu\cmn
 {\rm th}}^{\rm max}$ \quad & \quad $p_{_{\rm KS}}$ for $\log F_\nu$ \quad
 & \quad $p_{_{\rm KS}}$ for $\log E$ \quad
 & \quad $p_{_{\rm KS}}$ for $\rm DM_E$ \quad \\ \hline
 \ \ $\bf -0.6$ \ \ & {\bf 5.5} & {\bf 0.74}
 & {\bf 0.8266} & {\bf 0.6002} & {\bf 0.5234} \\ \hline
 $\bf -0.2$ & {\bf 2.7} & {\bf 0.70}
 & {\bf 0.5419} & {\bf 0.6350} & {\bf 0.5015} \\ \hline
 $-0.1$ & 2.3 & 0.76 & 0.6893 & 0.4480 & 0.5806 \\ \hline
 $-0.3$ & 3.2 & 0.81 & 0.6375 & 0.5079 & 0.4744 \\ \hline
 $-0.4$ & 3.4 & 0.80 & 0.6469 & 0.4042 & 0.5302 \\ \hline
 $-0.6$ & 5.2 & 0.79 & 0.5802 & 0.5028 & 0.5181 \\ \hline
 $-0.5$ & 6.5 & 0.79 & 0.7272 & 0.5815 & 0.5082 \\ \hline
 $-0.4$ & 7.3 & 0.77 & 0.6147 & 0.5132 & 0.5982 \\ \hline
 $-0.9$ & 8.0 & 0.70 & 0.9063 & 0.6155 & 0.5672 \\ \hline
 $0.1$ & 1.6 & 0.74 & 0.6627 & 0.4106 & 0.4457 \\ \hline
 \hline
 \end{tabular}
 \end{center}
 \vspace{-2mm}  
 \caption{\label{tab2} The same as in Table~\ref{tab1}, but for
 some examples of the acceptable CPL model.}
 \end{table}



 \begin{center}
 \begin{figure}[tb]
 \centering
 \vspace{0.5mm}  
 \includegraphics[width=0.82\textwidth]{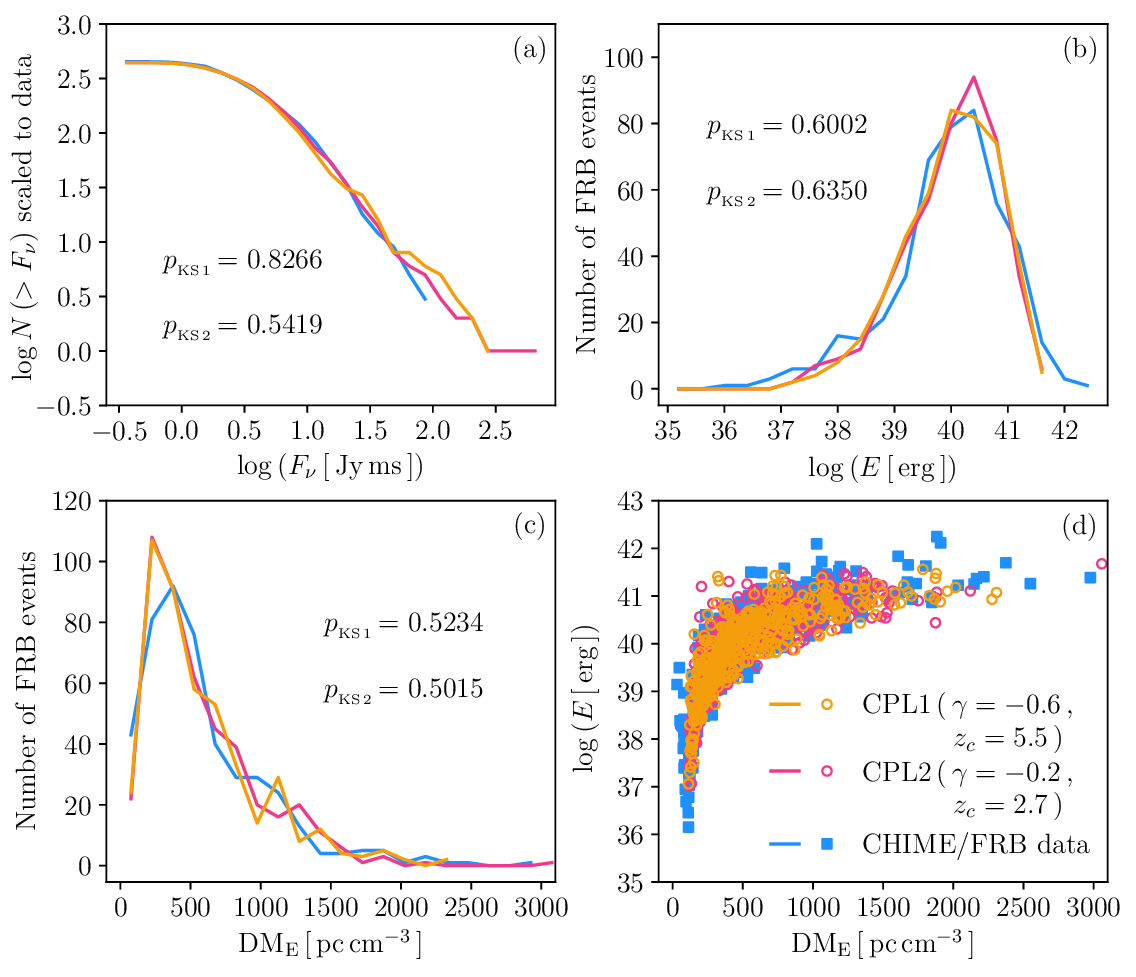}
 \vspace{-2mm}  
 \caption{\label{fig4} The same as in Fig.~\ref{fig2}, but for
 the CPL model with the model parameters given in panel~(d).}
 \end{figure}
 \end{center}


\vspace{-21.7mm} 

At first, we consider the SFH model in which the FRB distribution tracks SFH
 (namely ${\cal E}(z)\equiv 1$ in Eq.~(\ref{eq10}) while ${\rm SFH}(z)$ is
 given by Eq.~(\ref{eq5})). In this case, there is no free parameter in the
 FRB redshift distribution model. Noting that 2,000,000 mock FRBs are not
 enough to leave more than 300 FRBs filtered by the sensitivity threshold
 and instrumental selection effects, we have to generate 5,000,000 mock
 FRBs in the simulation. We scan the free parameter $\log F_{\nu\cmn {\rm
 th}}^{\rm max}$ in the sensitivity model from $-0.49$ to $2.0$ with step
 size $0.01$, and evaluate three p-values of the KS tests with respect to
 the $\log F_\nu$, $\log E$ and $\rm DM_E$ criteria for each value of $\log
 F_{\nu\cmn {\rm th}}^{\rm max}$. We find that the SFH model fails to
 reproduce all the $\log F_\nu$, $\log E$ and $\rm DM_E$ distributions of
 the CHIME/FRB sample, because for all values of $\log F_{\nu\cmn {\rm
 th}}^{\rm max}$, three $p_{_{\rm KS}}$ cannot be simultaneously larger than
 $0.1$. In Fig.~\ref{fig2}, we present a typical example with $\log F_{\nu
 \cmn {\rm th}}^{\rm max}=0.81$. Its KS p-values for both $\log E$ and
 $\rm DM_E$ criteria are ${\cal O}(10^{-4})\ll 0.05$. From panel~(d)
 of Fig.~\ref{fig2}, one can see that many mock FRBs with $\rm DM_E\gtrsim
 1000\;\dmunit$ are clearly outside the crowded region of the CHIME/FRB
 observational data. From panel~(c) of Fig.~\ref{fig2}, we find that the
 number of mock FRBs with relatively large ${\rm DM_E}\gtrsim 700\,\dmunit$
 is overproduced. This hints that a suppressed evolution at large $\rm
 DM_E$ (or equivalently high-redshifts) might be required by the CHIME/FRB
 observational data. Clearly, due to the very low p-values $p_{_{\rm KS}}
 \ll 0.05$, the CHIME/FRB observational data strongly reject the SFH model
 at very high confidence. We confirm the result for the SFH model found
 in~\cite{Zhang:2021kdu}. So, one should find some empirical
 FRB distribution models consistent with the CHIME/FRB observational data.


\begin{table}[tb]
 \renewcommand{\arraystretch}{1.7}
 \begin{center}
 \vspace{-8mm}  
 \begin{tabular}{ccccc} \hline\hline
 \quad $z_c$ \ \ &  \quad $\log F_{\nu\cmn {\rm th}}^{\rm max}$ \quad
 & \quad $p_{_{\rm KS}}$ for $\log F_\nu$ \quad
 & \quad $p_{_{\rm KS}}$ for $\log E$ \quad
 & \quad $p_{_{\rm KS}}$ for $\rm DM_E$ \quad \\ \hline
 {\bf 2.8} & {\bf 0.70} & {\bf 0.9614} & {\bf 0.2662} &
 {\bf 0.5479} \\ \hline
 2.9 & 0.77 & 0.9003 & 0.2869 & 0.4900 \\ \hline
 3.6 & 0.73 & 0.5305 & 0.2609 & 0.4994 \\ \hline
 2.4 & 0.72 & 0.8368 & 0.2101 & 0.3914 \\ \hline
 2.0 & 0.69 & 0.4177 & 0.2270 & 0.2526 \\ \hline
 \hline
 \end{tabular}
 \end{center}
 \vspace{-2mm}  
 \caption{\label{tab3} The same as in Table~\ref{tab1}, but for
 some examples of the acceptable CSFH model.}
 \end{table}



 \begin{center}
 \begin{figure}[tb]
 \centering
 \vspace{1mm}  
 \includegraphics[width=0.82\textwidth]{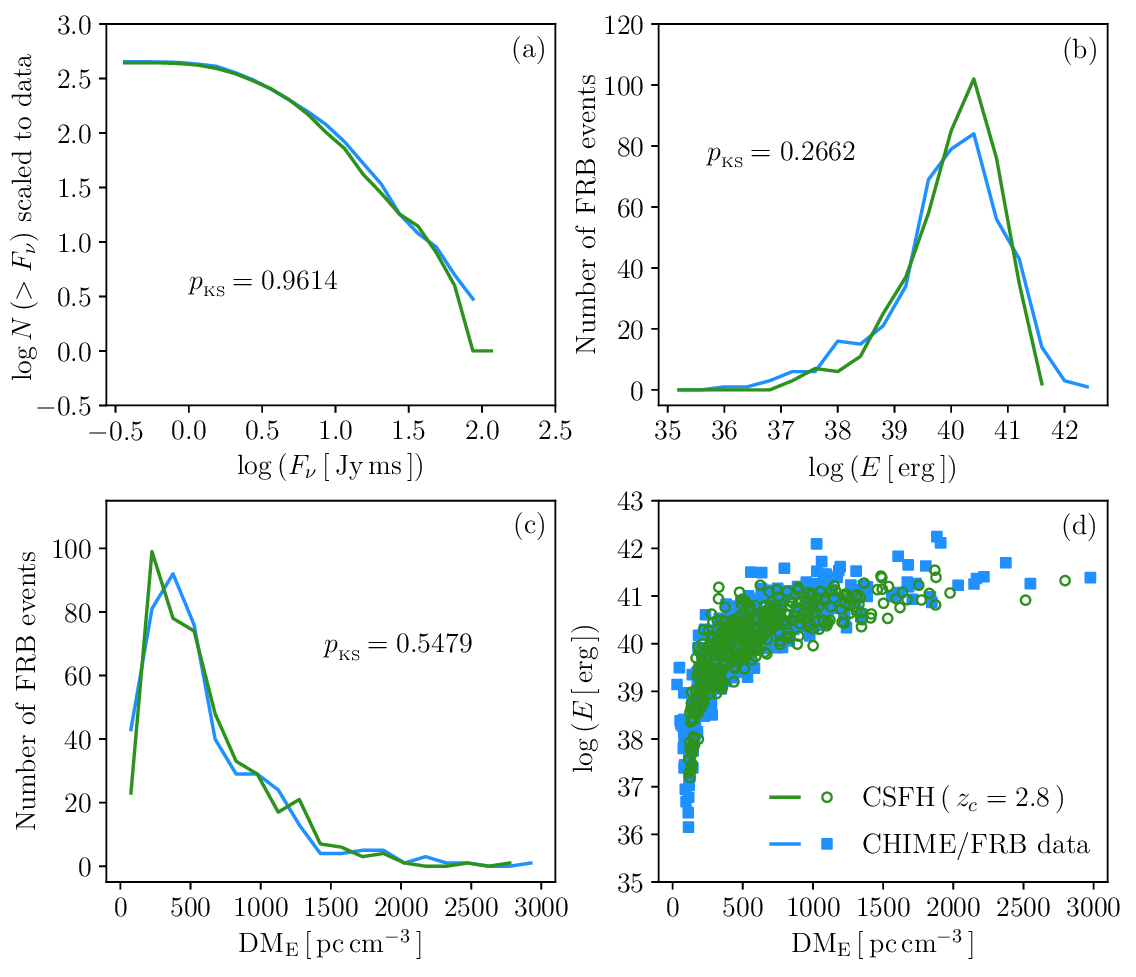}
 \caption{\label{fig5} The same as in Fig.~\ref{fig2}, but for
 the CSFH model with the model parameter given in panel~(d).}
 \end{figure}
 \end{center}


\vspace{-11mm} 

As mentioned above, motivated by the similar case of GRBs, we consider some
 SFH-based FRB redshift distribution models given by Eq.~(\ref{eq10}).
 Notice that for all the following models, it is enough to generate
 2,000,000 mock FRBs for each value combination of the free parameters in
 the redshift distribution model to leave more than 300 FRBs filtered by the
 sensitivity threshold and instrumental selection effects, comparable with
 the CHIME/FRB sample of 452 FRBs as mentioned in Sec.~\ref{sec2b}.

The simplest one is the PL model given by Eq.~(\ref{eq11}), which was
 also considered in the case of GRBs~\cite{Kistler:2007ud} (see also
 e.g.~\cite{Yuksel:2008cu}). There is only one free parameter $\gamma$ in
 the FRB redshift distribution model. We scan the free parameter $\gamma$
 from $-2.5$ to $+2.5$ with step size $0.1$, and scan $\log F_{\nu\cmn {\rm
 th}}^{\rm max}$ in the sensitivity model from $-0.49$ to $2.0$ with step
 size $0.01$. For each parameter combination $(\gamma,\,\log F_{\nu\cmn
 {\rm th}}^{\rm max})$, we evaluate three p-values of the KS tests with
 respect to the $\log F_\nu$, $\log E$ and $\rm DM_E$ criteria. Some
 examples are shown in Table~\ref{tab1}. We find that the PL model with some
 negative $\gamma$ cannot be rejected at high confidence by the CHIME/FRB
 observational data, with respect to all the three criteria for $\log
 F_\nu$, $\log E$ and $\rm DM_E$. In particular, we present an explicit
 example of the PL model with $\gamma=-1.1$ (and $\log F_{\nu\cmn {\rm
 th}}^{\rm max}=0.72$) in Fig.~\ref{fig3}, whose three p-values are
 all larger than $0.49$ simultaneously. So, the PL model with suitable
 parameters can be consistent with the CHIME/FRB observational data.
 Clearly, a negative $\gamma$ is commonly required. It is worth noting that
 ${\cal E}(z)\propto (1+z)^\gamma$ with a negative $\gamma$ for FRBs
 indicates a suppressed evolution with respect to SFH (n.b. Eq.~(\ref{eq10})
 with Eq.~(\ref{eq11})). On the contrary, as mentioned above, an enhanced
 evolution with respect to SFH (namely ${\cal E}(z)\propto (1+z)^{1.5}$) was
 found for GRBs~\cite{Kistler:2007ud} (see also e.g.~\cite{Yuksel:2008cu}).
 These two trends are opposite in fact. This suggests that FRBs might be not
 closely related to GRBs, and might shed light on the origin of FRBs.


\begin{table}[tb]
 \renewcommand{\arraystretch}{1.7}
 \begin{center}
 \vspace{-7mm}  
 \begin{tabular}{ccccccc} \hline\hline
 $\gamma_1$ & $\gamma_2$ & $s$ & \quad $\log F_{\nu\cmn {\rm th}}^{\rm max}$
 \quad & \quad $p_{_{\rm KS}}$ for $\log F_\nu$
 \quad & \quad $p_{_{\rm KS}}$ for $\log E$ \quad
 & \quad $p_{_{\rm KS}}$ for $\rm DM_E$ \quad \\ \hline
 \ \ $\bf -0.7$ \ \ & \ \ {\bf 1.1} \ \ & \ \ {\bf 1.9} \ \
 & {\bf 0.76} & {\bf 0.9825} & {\bf 0.7414} & {\bf 0.7609} \\ \hline
 $\bf -0.9$ & {\bf 0.6} & {\bf 0.4} & {\bf 0.77}
 & {\bf 0.8516} & {\bf 0.7044} & {\bf 0.7044} \\ \hline
 $-1.3$ & 0.6 & 0.4 & 0.75 & 0.8445 & 0.7149 & 0.7207 \\ \hline
 $-0.3$ & 1.4 & 0.7 & 0.88 & 0.7306 & 0.8934 & 0.7234 \\ \hline
 $-1.1$ & 0.4 & 0.7 & 0.77 & 0.7112 & 0.7576 & 0.7281 \\ \hline
 $-0.9$ & 0.6 & 2.4 & 0.72 & 0.9988 & 0.6401 & 0.6352 \\ \hline
 $-1.1$ & 0.8 & 3.2 & 0.76 & 0.9747 & 0.6659 & 0.6153 \\ \hline
 $-0.2$ & 2.2 & 3.6 & 0.76 & 0.9523 & 0.6461 & 0.6498 \\ \hline
 \hline
 \end{tabular}
 \end{center}
 \vspace{-2mm}  
 \caption{\label{tab4} The same as in Table~\ref{tab1}, but for
 some examples of the acceptable TSE model.}
 \end{table}



 \begin{center}
 \begin{figure}[tb]
 \centering
 \vspace{-11mm}  
 \includegraphics[width=0.82\textwidth]{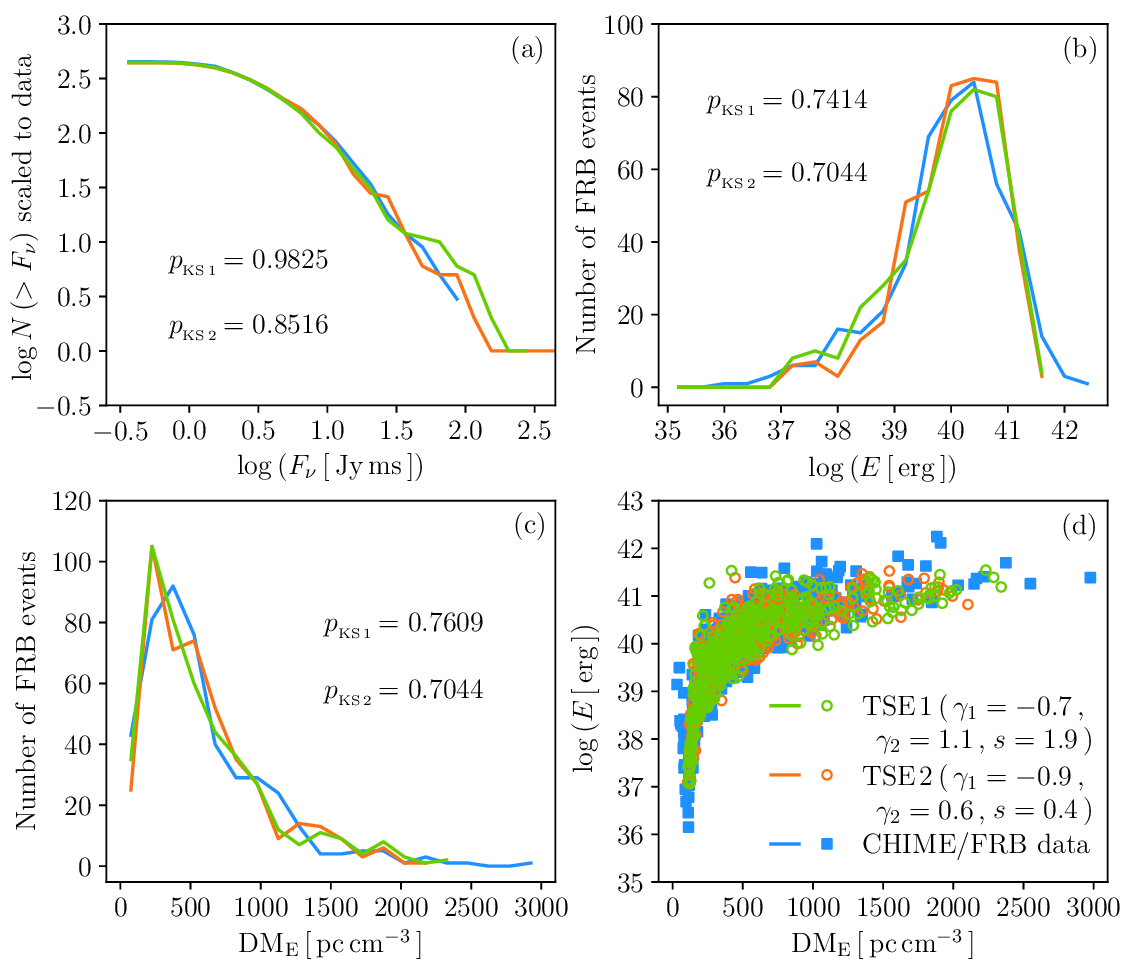}
 \caption{\label{fig6} The same as in Fig.~\ref{fig2}, but for
 the TSE model with the model parameters given in panel~(d).}
 \end{figure}
 \end{center}


\vspace{-10.7mm} 

Of course, an exponential cutoff might be taken into account as in the
 literature. So, it is reasonable to consider the CPL model given by
 Eq.~(\ref{eq12}), which has two free parameters $\gamma$ and $z_c\,$. It
 is expected that ${\cal E}(z)$ has a sharp cutoff around $z_c\,$. We scan
 the free parameters $\gamma$ from $-2.5$ to $+2.5$ with step size $0.1$,
 $z_c$ from $0.1$ to $8.0$ with step size $0.1$, and scan $\log F_{\nu\cmn
 {\rm th}}^{\rm max}$ in the sensitivity model from $-0.49$ to $2.0$ with
 step size $0.01$. For each parameter combination $(\gamma\cmn z_c\,\cmn
 \log F_{\nu\cmn {\rm th}}^{\rm max})$, we evaluate three p-values of the KS
 tests with respect to the $\log F_\nu$, $\log E$ and $\rm DM_E$ criteria.
 Some examples are shown in Table~\ref{tab2}. We find that the CPL model
 with suitable parameters cannot be rejected by the CHIME/FRB observational
 data, with respect to all the three criteria for $\log F_\nu$, $\log E$ and
 $\rm DM_E$. For a very high $z_c$ (say, $>7$), the CPL model effectively
 reduces to the PL model, since there are only a few FRBs at very high
 redshifts. As shown in Table~\ref{tab2}, some moderate $z_c$ are still
 viable, and in these cases their corresponding $\gamma$ need not to be very
 negative. In Fig.~\ref{fig4}, we present two explicit examples of the CPL
 model with different model parameters. Obviously, the CPL model with
 suitable parameters can also be consistent with the CHIME/FRB observational
 data. For many parameters, three p-values for the $\log F_\nu$, $\log
 E$ and $\rm DM_E$ criteria are all larger than $0.5$ simultaneously.
 Comparing three $p_{_{\rm KS}}$ in Table~\ref{tab2} with the ones in
 Table~\ref{tab1}, the CPL model is notably favored over the PL model, since
 all the three p-values for the $\log F_\nu$, $\log E$ and $\rm DM_E$
 criteria are closer to $1$, especially for both $\log E$ and $\rm DM_E$
 criteria. Note that a negative $\gamma$ in the CPL model is
 commonly required by the CHIME/FRB observational data, which indicates a
 suppressed evolution with respect to SFH again.

A special case of the CPL model is of interest. If $\gamma=0$ in the CPL
 model, it becomes SFH with an exponential cutoff (CSFH) model. It might
 loosely track SFH at low-redshifts if the sharp cutoff appears around a
 fairly high $z_c\,$. We scan the only free parameter $z_c$ in ${\cal E}(z)$
 from $0.1$ to $8.0$ with step size $0.1$, and scan $\log F_{\nu\cmn {\rm
 th}}^{\rm max}$ in the sensitivity model from $-0.49$ to $2.0$ with step
 size $0.01$. For each parameter combination $(z_c\,\cmn\log F_{\nu\cmn
 {\rm th}}^{\rm max})$, we evaluate three p-values of the KS tests with
 respect to the $\log F_\nu$, $\log E$ and $\rm DM_E$ criteria. As shown
 in Table~\ref{tab3} and Fig.~\ref{fig5}, the CSFH model with suitable
 parameters cannot be rejected at high confidence by the
 CHIME/FRB observational data, with respect to all the three criteria for
 $\log F_\nu$, $\log E$ and $\rm DM_E$. This might be welcome to some
 theoretical models for the FRB engines, in which FRBs still track SFH
 at low-redshifts, only at the price of breaking down at fairly
 high-redshifts, where some unknown factors necessary to trigger FRBs
 might be not ready.

In the TSE model given by Eq.~(\ref{eq13}), due to the two-segment behavior
 of ${\cal E}(z)$, the evolution with respect to SFH might be enhanced
 (suppressed) at low-redshifts, and suppressed (enhanced) at high-redshifts,
 respectively. Of course, it is also possible to be suppressed (enhanced) at
 both low- and high-redshifts but with different power law indices. There
 are three free parameters $\gamma_1$, $\gamma_2$ and $s$ in ${\cal E}(z)$.
 We scan $\gamma_1$ from $-2.0$ to $+2.0$, $\gamma_2$ from $-3.0$ to $+3.0$,
 $s$ from $0.1$ to $4.0$. To save the computational power and time, we
 firstly scan them with a rough step size $0.2$, and then re-scan them with
 a finer step size $0.1$ in the parameter spaces having high p-values. For
 each parameter combination $(\gamma_1\cmn\gamma_2\cmn s)$, we scan $\log
 F_{\nu\cmn {\rm th}}^{\rm max}$ in the sensitivity model from $-0.49$ to
 $2.0$ with step size $0.01$ and evaluate three p-values of the KS tests
 with respect to the $\log F_\nu$, $\log E$ and $\rm DM_E$ criteria. Some
 examples are shown in Table~\ref{tab4}. We find that the TSE model with
 suitable parameters cannot be rejected by the CHIME/FRB observational
 data, with respect to all the three criteria for $\log F_\nu$, $\log E$
 and $\rm DM_E$. As shown in Table~\ref{tab4}, for many parameters, three
 p-values for the $\log F_\nu$, $\log E$ and $\rm DM_E$ criteria are all
 larger than $0.7$ simultaneously. In fact, there are more examples with
 three p-values all larger than $0.6$ simultaneously, and we have not
 clearly shown them in Table~\ref{tab4}. In Fig.~\ref{fig6}, we present
 two explicit examples of the TSE model with different model parameters.
 Obviously, the TSE model with suitable parameters can be fully consistent
 with the CHIME/FRB observational data. Comparing three $p_{_{\rm KS}}$ in
 Table~\ref{tab4} with the ones in Table~\ref{tab2}, the TSE model is
 notably favored over the CPL model, since all the three p-values for the
 $\log F_\nu$, $\log E$ and $\rm DM_E$ criteria are closer to $1$. Note
 that a negative $\gamma_1$ and a positive $\gamma_2$ in the TSE model are
 simultaneously favored by the CHIME/FRB observational data, which indicate
 a suppressed evolution with respect to SFH at both low- and high-redshifts.

\begin{center}
 \begin{figure}[tb]
 \centering
 \vspace{-11mm}  
 \includegraphics[width=0.82\textwidth]{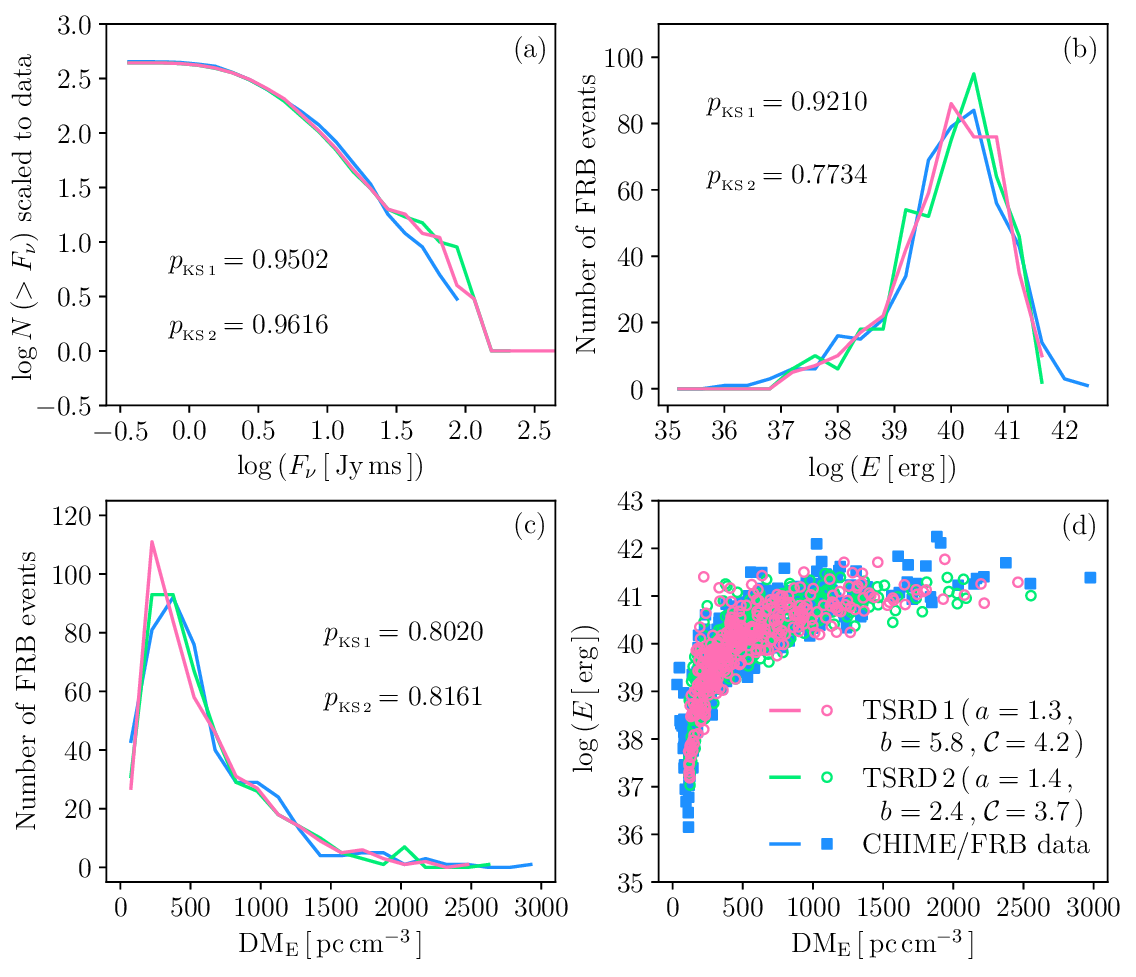}
 \caption{\label{fig7} The same as in Fig.~\ref{fig2}, but for
 the TSRD model with the model parameters given in panel~(d).}
 \end{figure}
 \end{center}

Since ${\rm SFH}(z)$ in Eq.~(\ref{eq5}) has also a two-segment behavior,
 and ${\cal E}(z)$ in Eq.~(\ref{eq10}) suppresses (enhances) it with
 different power law indices at low- and high-redshifts, we can rather
 parameterize $dN/(dt\,dV)$ directly by using a single two-segment function
 given by Eq.~(\ref{eq14}). There are three free parameters $a$, $b$ and
 $\cal C$ in this non-SFH-based TSRD model. We scan $a$ from $-2.0$ to
 $+3.0$, $b$ from $-3.0$ to $+6.0$, $\cal C$ from $0.1$ to $5.0$. To save
 the computational power and time, similar to the TSE model, we firstly scan
 them with a rough step size $0.2$, and then re-scan them with a finer step
 size $0.1$ in the parameter spaces having high p-values. For each parameter
 combination $(a\cmn b\cmn {\cal C})$, we scan $\log F_{\nu\cmn {\rm
 th}}^{\rm max}$ in the sensitivity model from $-0.49$ to $2.0$ with step
 size $0.01$ and evaluate three p-values of the KS tests with respect to
 the $\log F_\nu$, $\log E$ and $\rm DM_E$ criteria. Some examples are shown
 in Table~\ref{tab5}. For many parameters, three p-values for the $\log
 F_\nu$, $\log E$ and $\rm DM_E$ criteria are all larger than $0.75$
 simultaneously (all $>0.8$ for a few parameters). In Fig.~\ref{fig7}, we
 present two explicit examples of the TSRD model with different model
 parameters. Clearly, the TSRD model with suitable parameters can be fully
 consistent with the CHIME/FRB observational data. Comparing three $p_{_{\rm
 KS}}$ in Table~\ref{tab5} with the ones in Table~\ref{tab4}, the TSRD model
 is slightly favored over the TSE model, since all the three p-values for
 the $\log F_\nu$, $\log E$ and $\rm DM_E$ criteria are closer to $1$.
 For the most favored TSRD parameters $a=1.3$, $b=5.8$ and ${\cal C}=4.2$
 in Table~\ref{tab5}, $dN/(dt\,dV)$ in Eq.~(\ref{eq14}) behaves like
 $(1+z)^{1.3}$ and $(1+z)^{-5.8}$ at low- and high-redshifts, respectively.
 Noting that ${\rm SFH}(z)$ in Eq.~(\ref{eq5}) behaves like $(1+z)^{2.6}$
 and $(1+z)^{-3.6}$ at low- and high-redshifts respectively, the most
 favored TSRD model in Table~\ref{tab5} is suppressed with respect to SFH
 at both low- and high-redshifts. However, since the TSRD model is not based
 on SFH, it is hard to directly compare the TSRD model with the SFH model in
 general. Although all $p_{_{\rm KS}}>0.75$ examples in Table~\ref{tab5}
 have $b>a>0$, there are still many acceptable cases with quite different
 parameters, e.g. both negative $a$ and $b$ (which are not presented
 in Table~\ref{tab5}), but their three $p_{_{\rm KS}}>0.6$ or $0.5$
 simultaneously. For those parameters, the TSRD model has different
 behaviors. In fact, the CHIME/FRB observational data can be consistent with
 quite different TSRD models. A much larger observational sample is required
 to differentiate them.

In brief, we have tested six types of FRB redshift distribution models,
 and found that many of them can be fully consistent with the CHIME/FRB
 observational data for some suitable model parameters. According to
 three KS p-values for the $\log F_\nu$, $\log E$ and $\rm DM_E$ criteria,
 they are favored in the following order: TSRD $\gtrsim$ TSE $>$ CPL $>$
 PL $>$ CSFH $\gg$ SFH. Noting that they have 3, 3, 2, 1, 1, 0 free
 parameters in the redshift distribution models respectively, this order
 is reasonable in fact.


\begin{table}[tb]
 \renewcommand{\arraystretch}{1.7}
 \begin{center}
 \vspace{-9mm}  
 \begin{tabular}{ccccccc} \hline\hline
 $a$ & $b$ & $\cal C$ & \quad $\log F_{\nu\cmn {\rm th}}^{\rm max}$
 \quad & \quad $p_{_{\rm KS}}$ for $\log F_\nu$
 \quad & \quad $p_{_{\rm KS}}$ for $\log E$ \quad
 & \quad $p_{_{\rm KS}}$ for $\rm DM_E$ \quad \\ \hline
 \ \ {\bf 1.3} \ \ & \ \ {\bf 5.8} \ \ & \ \ {\bf 4.2} \ \ & {\bf 0.71}
 & {\bf 0.9502} & {\bf 0.9210} & {\bf 0.8020} \\ \hline
 {\bf 1.4} & {\bf 2.4} & {\bf 3.7} & {\bf 0.66}
 & {\bf 0.9616} & {\bf 0.7734} & {\bf 0.8161} \\ \hline
 1.1 & 2.7 & 4.3 & 0.72 & 0.7670 & 0.8397 & 0.7874 \\ \hline
 0.9 & 3.5 & 4.8 & 0.72 & 0.7748 & 0.7801 & 0.8532 \\ \hline
 1.4 & 2.8 & 4.1 & 0.61 & 0.9215 & 0.7611 & 0.7717 \\ \hline
 1.3 & 2.3 & 5.0 & 0.56 & 0.8841 & 0.8288 & 0.7174 \\ \hline
 1.6 & 1.3 & 2.8 & 0.69 & 0.8816 & 0.7396 & 0.7722 \\ \hline
 1.2 & 4.7 & 4.1 & 0.72 & 0.8210 & 0.7945 & 0.7174 \\ \hline
 \hline
 \end{tabular}
 \end{center}
 \vspace{-2mm}  
 \caption{\label{tab5} The same as in Table~\ref{tab1}, but for
 some examples of the acceptable TSRD model.}
 \end{table}



\vspace{-11mm} 


\section{Concluding remarks}\label{sec4}

Nowadays, FRBs have become one of the active fields in astronomy and
 cosmology. However, the origin of FRBs is still unknown to date. The
 studies on the intrinsic FRB distributions might help us to reveal the
 possible origins of FRBs, and improve the simulations for FRB cosmology.
 Recently, the first CHIME/FRB catalog of 536 events was released. Such
 a large uniform sample of FRBs detected by a single telescope is very
 valuable to test the FRB distributions. Later, it has been claimed
 that the FRB distribution model tracking SFH was rejected by the first
 CHIME/FRB catalog. In the present work, we consider some empirical FRB
 distribution models, and find that many of them can be fully consistent
 with the CHIME/FRB observational data for some suitable model parameters.

Some remarks are in order. At first, as shown in this work, we
 independently test the FRB distribution model tracking SFH against the
 first CHIME/FRB observational data, and affirmatively find that it can be
 rejected at very high confidence. We confirm this important result found
 in~\cite{Zhang:2021kdu}. It clearly means that other FRB distribution
 models are required by the observational data.

In most of the models considered here, a suppressed evolution with respect
 to SFH is found for FRBs. A negative power law index $\gamma$ in the PL
 and CPL models is commonly required by the CHIME/FRB observational data.
 An exponential cutoff in the CPL and CSFH models is actually a sharp
 suppression. A negative $\gamma_1$ and a positive $\gamma_2$ in the TSE
 model are simultaneously favored by the CHIME/FRB observational data,
 which also indicate a suppressed evolution with respect to SFH for FRBs.
 The most favored TSRD model is also suppressed with respect to SFH, as
 mentioned above. On the contrary, an enhanced evolution with respect to
 SFH was found for GRBs~\cite{Kistler:2007ud} (see also
 e.g.~\cite{Yuksel:2008cu}). This suggests that FRBs might be not closely
 related to GRBs, and might shed light on the origin of FRBs.

On the other hand, the CSFH model with suitable parameters cannot be
 rejected at high confidence by the CHIME/FRB observational data, although
 the corresponding KS p-values are not so high. This might be welcome to
 some theoretical models for the FRB engines, in which FRBs still track SFH
 at low-redshifts, only at the price of breaking down at fairly
 high-redshifts, where some unknown factors necessary to trigger FRBs
 might be not ready.

In fact, all the ``\,successful\,'' models mentioned above effectively
 require a certain degree of ``\,delay\,'' with respect to SFH. This point
 can be clearly seen from Fig.~\ref{fig1}. The peaks of these distribution
 models are effectively shifted to the low-redshifts. In some sense, our
 phenomenological models are in general consistent with the conclusion
 of~\cite{Zhang:2021kdu}, namely a certain degree of ``\,delay\,'' with
 respect to SFH is required by the observational data.

The main difference between the present work and~\cite{Zhang:2021kdu} is
 that the latter claimed a much more delayed population than ours. Note
 that a three-segment empirical SFH model from~\cite{Yuksel:2008cu}
 (which also includes GRB contribution from high-redshifts) was used
 in~\cite{Zhang:2021kdu}, while we instead use the two-segment empirical
 SFH model from~\cite{Madau:2016jbv} (see also~\cite{Madau:2014bja}) in
 the present work. This might be the main reason for the much more delay
 required by~\cite{Zhang:2021kdu} to match the same data.

It is worth noting that our FRB distribution models are mainly empirical
 or phenomenological ones. The physical mechanisms behind them are not
 clear. In the case of GRBs, an enhanced evolution with respect to SFH was
 well motivated, namely stars formed at higher redshifts tend to have
 lower metallicity which would facilitate generating more long GRBs (see
 e.g.~\cite{Kistler:2007ud,Virgili:2011jc,Wang:2014nna}). However, the
 trend is opposite in the case of FRBs. To date, around 20 FRB host galaxies
 were identified~\cite{Heintz:2020}, whose metallicities are in the range
 from 8.08 to 9.03 (most of them are larger than 8.7, the metallicity upper
 cutoff for long GRBs \cite{Wang:2014nna}). It seems that higher metallicity
 might facilitate generating more FRBs or their (unknown) progenitors
 (but correlation does not mean causality). As mentioned above, all the
 ``\,successful\,'' models effectively require a certain degree
 of ``\,delay\,'' with respect to SFH. Such kind of delay might rise from
 various physical reasons. One of the physically motivated models might be
 the compact binary merger models considered in e.g.~\cite{Zhang:2020ass,
 Zhang:2021kdu,Cao:2018yzp,Locatelli:2018anc}. A binary system must undergo
 a long inspiral phase before the final merger, and hence a significant
 delay is necessary~\cite{Zhang:2020ass,Zhang:2021kdu}. On the other hand,
 the models invoking AGNs, white dwarfs, cosmic comb, black holes, cosmic
 strings, or interactions between asteroids/comets and neutron stars,
 also require a delay less than the one of compact binary merger models
 \cite{Zhang:2020ass,Zhang:2021kdu}. These FRB models that involve a
 delay with respect to SFH might be regarded as the physical motivations
 for our distribution models considered here.

In this work, for each value combination of the free parameters in the
 redshift distribution model, we generate 2,000,000 mock FRBs (or 5,000,000
 mock FRBs for the SFH model) to ensure that there are still enough FRBs
 filtered by the sensitivity threshold and instrumental selection effects
 (say, $N>300$), comparable with the CHIME/FRB sample consisting of 452 FRBs
 as mentioned in Sec.~\ref{sec2b}. For example, in the CPL model having two
 free model parameters, we scan the free parameters $\gamma$ from $-2.5$ to
 $+2.5$ with step size $0.1$, $z_c$ from $0.1$ to $8.0$ with step size $0.1$,
 and hence there are about $50\times 80=4000$ value combinations of
 $(\gamma\cmn z_c)$. Thus, we should generate about $4000\times 2,000,000=
 8,000,000,000$ mock FRBs in total. For the TSE and TSRD models having three
 free model parameters, the cost of computation increases dramatically, as
 expected. Clearly, this consumes a large amount of computational power and
 time. Therefore, we have to scan the model parameters roughly (with a step
 size $0.1$ commonly) in the relatively narrow parameter spaces (especially
 for the TSE and TSRD models) to save the computational power and time. In
 this way, we might miss the optimal parameters with maximum p-values for
 the KS tests. But at least we do show that the FRB distribution models
 considered in this work can be fully consistent with the CHIME/FRB
 observational data for some explicit parameters. Of course, if we could
 use some significantly powerful computing resources to this end, it is
 expected that the best model parameters will be precisely found. We hope
 us or someone else can do that eventually in the future.


\section*{ACKNOWLEDGEMENTS}

We thank the anonymous referee for quite useful comments and suggestions,
 which helped us to improve this work. We are grateful to Hua-Kai~Deng,
 Han-Yue~Guo, Shupeng~Song, Zhong-Xi~Yu and Jing-Yi~Jia for kind help and
 useful discussions. This work was supported in part by NSFC under Grants
 No.~11975046 and No.~11575022.

\renewcommand{\baselinestretch}{1.1}


\end{document}